\documentclass[grl]{agu2001}
 \usepackage[dvips]{graphicx}
\lefthead{Grasso and Zaliapin}
\righthead{Volcano predictability} 
 \authoraddr{Jean-Robert Grasso, LGIT, Observatoire de Grenoble, BP 53X,
38041  Grenoble Cedex, France. (Jean-Robert.Grasso@obs.ujf-grenoble.fr)}
\authoraddr{Ilya Zaliapin,  Institute of Geophysics and Planetary Physics,
University of California, Los Angeles, California, USA and
International Institute for Earthquake Prediction Theory and
Mathematical Geophysics, Russian Ac. Sci., Moscow, RUSSIA (zal@ess.ucla.edu)}
\input{update.tex}
\begin{document}

\title{Predictability of Volcano Eruption:
lessons from a basaltic effusive volcano.}

\author{Jean-Robert Grasso}

\affil{Laboratoire de G{\'e}ophysique
Interne et Tectonophysique, Observatoire de Grenoble, France
and Institute of Geophysics and Planetary Physics,
University of California, Los Angeles, California, USA}

\author{Ilya Zaliapin}

\affil{Institute of Geophysics and Planetary Physics,
University of California, Los Angeles, California, USA and
International Institute for Earthquake Prediction Theory and
Mathematical Geophysics, Russian Ac. Sci., Moscow, RUSSIA}

\newcommand{\be}{\begin{equation}}
\newcommand{\ee}{\end{equation}}
\newcommand{\ba}{\begin{eqnarray}}
\newcommand{\ea}{\end{eqnarray}}
\newenvironment{technical}{\begin{quotation}\small}{\end{quotation}}

\begin{abstract}

 Volcano eruption forecast remains a challenging and
 controversial problem despite the fact that data from
 volcano monitoring significantly increased in quantity
 and quality during the last decades.
 This study uses pattern recognition techniques to quantify
 the predictability of the 15 Piton de la Fournaise (PdlF)
 eruptions in the 1988-2001 period using increase of the
 daily seismicity rate as a precursor.
 Lead time of this prediction is a few days to weeks.
 Using the daily seismicity rate, we formulate a simple
  prediction rule, use it for
 retrospective prediction of the 15 eruptions,
 and test the prediction quality with error diagrams.
 The best prediction performance corresponds to averaging the
 daily seismicity rate over 5 days and issuing a prediction
 alarm for 5 days.
 65\% of the eruptions are predicted for an alarm
 duration less than 20\% of the time considered.
 Even though this result is concomitant of a large number of
 false alarms, it is obtained with a crude counting of daily
 events that are available from most volcano observatories.

\end{abstract}

\begin{article}

\section{Introduction}

The effective prediction success of volcanic eruptions
is rare if one defines ``prediction'' as a precise statement
of time, place, and ideally the nature and size of an
impending activity
[{\it Minakami}, 1960; {\it Swanson et al.}, 1985; {\it Voight}, 1988;
{\it Tilling and Lipman}, 1993; {\it Chouet}, 1996; {\it Mcnutt}, 1996].
A noteworthy obstacle is that most studies do not quantify the
effectiveness and reliability of proposed predictions, and often
do not surpass the analysis of a unique success on a single
case history with the lack of systematic description of
forecasting results.
In this study we focus on rigorous quantification of the
predictive power of the increase in the daily seismicity rate
--- a well-known and probably the simplest volcano
premonitory pattern.

Following {\it Minakami} [1960], {\it Kagan and Knopoff} [1987],
{\it Keilis-Borok} [2002],
we do not consider here deterministic predictions, and define a prediction
to be
``a formal rule whereby the available observable manifold of eruption
occurrence is significantly contracted and for this contracted manifold a
probability of occurrence of an eruption is significantly increased''
[{\it Kagan and Knopoff}, 1987].
To quantify the effectiveness and reliability of such predictions
we use error diagrams
[{\it Kagan and Knopoff}, 1987; {\it Molchan}, 1997].

Previous attempts in probalistic forecast of volcanic eruptions
used seismicity data in combination with other observations or alone
 [{\it Minakami}, 1960; {\it Klein}, 1984;  {\it Mulargia et al.}, 1991, 1992].
These studies did not quantify the
prediction schemes in the error diagram framework.
{\it Minakami} [1960] was a pioneer in the development of seismic statistics method
for volcano monitoring. Based on the data from the andesitic Asama volcano,
Honshu, he uses the increase in five-day frequencies of earthquakes to derive
an increase in the probability for an eruption
in the next 5 days.
{\it Klein} [1984] tests the precursory significance of geodetic data,
daily seismicity rate, and tides before the 29 eruptions during 1959-1979
at the Kilauea volcano, Hawaii.
He derives a probabilistic prediction scheme that applies for eruptions
anywhere on the volcano and can give 1- or 30-days forecast.
The forecasting ability of daily seismicity rate is shown to be
better than random at 90\% confidence in forecasts on the time scale
of 1 or 30 days using small earthquakes that occur in the caldera.
A better performance is achieved with a 99\% confidence when using
located earthquakes only, in forecasts on the time scale of 1 day.
{\it Mulargia et al.} [1991, 1992] use regional seimicity to define clusters
of seismic events within 120 km distance of Etna volcano.
Clusters within this regional seismicity are found
within 40 days before 9 out of 11 flank eruptions in the
1974-1989 period.
On the same period no statistically significant patterns are identified
40 days before and after the 10 summit eruptions.

As a test site we choose the PdlF volcano, the most active volcano
worldwide for the last decades with 15 eruptions in the 1988-2001 period.
On this site the volcanic risk remains low because most of
the eruptions are effusive and occurred in an area that is not inhabited.
For the PdlF site, the increase in seismicity rate and an increase
in deformation rate have been reported within a few hours prior to an
eruption
(e.g. [{\it Lenat et al.}, 1989; {\it Grasso and Bachelery}, 1995;
{\it Sapin et al.}, 1996; {\it Aki and Ferrazzini}, 2000;
{\it Collombet et al.}, 2003; {\it Lenat et al.}, 1989;
{\it Cayol and Cornet}, 1998].
Although the deformation data are very efficient to locate the lava
outflow vents from a few hours to minutes before the surface lava flow,
there is not yet a long term catalog available to test how they
can be used to forecast an eruption days to weeks in advance.

In this study we quantify the predictability of the PdlF eruptions
on the longer time scale of a {\it few days} to {\it weeks} prior to an
eruption.
{\it Collombet et al.} [2003] show that accelerating seismicity rate
weeks prior to the PdlF eruptions can be recovered on average using the
superposed epoch analysis before numerous eruptions.
Here we show that the increase of the daily seismicity rate
is useful as well to forecast individual eruptions.
This is achieved by rigorous quantification of
the prediction performance by introducing error diagrams
[{\it Kagan and Knopoff,} 1987; {\it Molchan,} 1997]
to choose among competitive prediction strategies.

\section{Data}

The PdlF hot spot volcano is a shield volcano with an effusive erupting
style due to low viscosity basaltic magma.
During 1988-2001 period the seismicity at the PdlF site remained low,
with $M_{\rm max} =3.5$, and was localized within a radius of a few
km beneath the central caldera.
Less than 10\% of these small events are located, most of them being
only recorded by the three summit stations located 3 km apart from each other.
Contrary to the Mauna Loa - Kilauea volcanic system,
there is no seismically active flank sliding or basal faulting on the PdlF.
Contrary to the Etna volcano, there is no tectonic interaction with
neighboring active structures.
Accordingly, the PdlF seismicity is one of the best candidates to be
purely driven by volcano dynamics.
This seismogenic volume is also thought to be
the main path for the magma to flow from a shallow storage system
toward the surface [{\it Lenat and Bachelery}, 1990;
{\it Sapin et al.}, 1996; {\it Bachelery}, 1999; {\it Aki and Ferrazini}, 2000].

The PdlF seismicity catalog consists of data from the 16 seismic stations
[{\it Sapin et al.}, 1996; {\it Aki and Ferrazzini}, 2000].
During the May 1988- June 2001 period the geometry and instrumental
characteristics of the seismic network remained stable,
 with a magnitude detection threshold of 0.5
[{\it Collombet et al.}, 2003].
In this period 15 eruptions were seismically monitored.
We use here the seismicity rate of the volcano tectonic (VT) events,
excluding long period (LP) events or rockfall signals.
The number of LP events at the PdlF site is insignificant compared
to the number of VT events.
For example, the eruption of 1998 was acconpanied by a single LP
event 4 hours before the surface lava flow
[{\it Aki and Ferrazzini}, 2000], and 2500 VT events had been
recorded at that time.

\section{Synthesis of seismicity pattern before eruptions}

Although the peaks of seismicity rate clearly correlate with eruption
days (see Figure~1 in [{\it Collombet et al}, 2003]), it is difficult
to identify a long-term seismicity pattern before each eruption,
except possibly during the last few hours before surface lava flow
[{\it Lenat et al.}, 1989; {\it Sapin et al.}, 1996;
{\it Aki and Ferrazzini}, 2000; {\it Collombet et al}, 2003].
For all the 15 PdlF eruptions the hourly seismicity rate during
the seismicity crisis that precedes each surface lava flow
is roughly constant with values ranging from 60 to 300 events/hr,
with an average value  of 120 events/hr.
The average crisis duration is 4 hrs, the extreme values
ranging from 0.5 hours for the may 1988 eruption to 36 hours for the 1998 eruption.
No correlation is found between the seismic rates or the durations
of the crisis and the erupted volumes.
Because there is no recurrent migration of seismicity during these crises
[e.g. {\it Sapin et al.}, 1996] we suggested, as proposed by
{\it Rubin et al.} [1998], that damage is neither
directly related to the dyke tip, nor does it always map the dyke propagation.
It is the response to dike intrusion of parts of the volcano
edifice that are close to failure [e.g. {\it Grasso and Bachelery}, 1995].

\begin{figure}
\resizebox{7.5cm}{5cm}{\includegraphics[width=7cm]{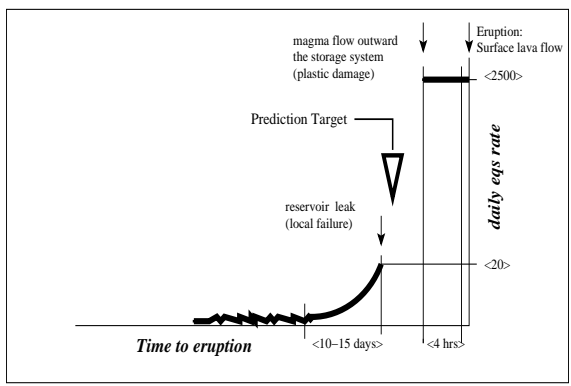}}
\caption{Average pre-eruptive pattern before a PdlF eruption.
This behavior is obtained by
averaging the seismicity rate over the 15 eruptions during
1988-2001.
}
\label{rfig2}
\end{figure}

We synthesize the pre-eruption seismicity rate on the PdlF volcano
as a 3 step process (Figure \ref{rfig2}).
First, the seismicity rate increases in average and
it follows a power law
 10-15 days prior the eruption [{\it Collombet et al}, 2003].
This is reminiscent of the average foreshock patterns observed for earthquakes
[{\it Jones and Molnar}, 1979; {\it Helmstetter and Sornette}, 2003].
As for earthquakes, we suggest that this pattern illuminates a local
damage process rather than a macroscopic failure, the damage being localized
within the magma storage system a few km below the volcano
[e.g. {\it Sapin et al.}, 1996].
This average acceleration is different from the acceleration proposed
prior to each single eruption by {\it Voight} [1988],
or individual large earthquakes [e.g. {\it Bufe and Varnes}, 1993].
The second phase is seismically mapped by a discontinuity in seismicity
rate from a peak value $<20$ events/day to a $>2000$ events/day
constant rate (Figure \ref{rfig2}).
We suggest it corresponds to the onset of the magma flow outward of
the storage system.
The third phase is characterized by a constant strong seismicity
rate during each crisis.
We suggest it corresponds to the damage induced by fluid flow, either
as a diffuse response to dyke propagation in an heterogeneous rock
matrix or as damage in the open reservoir walls during fluid flow.

This pre-eruption scheme helps both to clarify the eruption phases on
the PdlF and to define our prediction targets.
If one uses a conventional definition of the target as the onset
time of surface lava flow, then all the eruptions
can be predicted a few hours in advance by choosing a daily
seismicity rate larger than 60 events/day as an alarm threshold.
For this threshold value the seismic crisis that did not
end up in an eruption are false alarms.
They are post-labelled at the observatory as intrusion,
and are part of the endogeneous growth of any volcano.
We aim to find precursory patterns before the outward
magma flow from the reservoir system.
Accordingly, we define our target as the onset of a reservoir leak as
mapped by the end of the average acceleration process and before the
onset of the eruption crisis  (Figure \ref{rfig2}).
This target possibly maps a local failure in the reservoir walls,
contemporary to the onset of outward magma flow from the reservoir,
and corresponds to predicting eruptions more than one day in advance.
Thus, our problem is different from that posed by {\it Klein} [1984].

\section{Prediction scheme and error diagram}

Here we follow a pattern recognition approach [e.g. {\it Gelfand et al.},
 1976] to predict rare extreme events in complex systems;
this approach is reviewed by {\it Keilis-Borok } [2002].
To use pattern recognition techniques as a forecasting tool we
define 3 steps in the data analysis.
First we consider a sequence of VT earthquake occurrence
times
${\it C}= {\it t_e:~e=1,2,\dots E; ~t_e\le t_{e+1}}.$
Note that we use neither magnitude nor location of events.
Second, on the sequence $C$ we define a function
$N(t,s)$ as the number of earthquakes within the time
window $[t-s,~t]$, $s$ being a numerical parameter.
This functional is calculated for the time interval
considered with different values of numerical parameter $s$.
Third, an {\it alarm} is triggered when the functional
$N(t,s)$ exceeds a predefined threshold $N_0$.
The threshold $N_0$ is usually chosen as a certain percentile of
the distribution function for the functional $N(t,s)$.
The alarm is declared for a time interval $\Delta$.
The alarm is terminated after an eruption occurs or the
time $\Delta$ expires, whichever comes first.
Our prediction scheme depends on three parameters:
time window $s$, threshold $N_0$,
and duration $\Delta$ of alarms.
The quality of this kind of prediction is evaluated
with help of  "error diagrams"
which are a key element in evaluating
a prediction algorithm
[{\it Kagan and Knopoff,} 1987;
{\it Molchan,} 1997].

\begin{figure}
\resizebox{7cm}{4cm}{ \includegraphics[width=7cm]{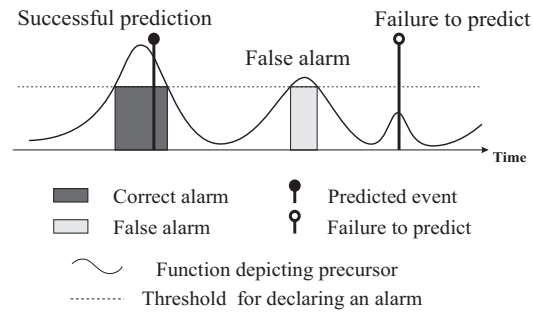}}
\caption{Prediction scheme and prediction outcomes.
}
\label{rfig3}
\end{figure}

The definition of an error diagram is the following.
Consider prediction by the scheme described above.
We continously monitor seismicity, declare alarms
when the functional $N(t,s)$ exceeds the threshold,
and count the prediction outcomes (Figure \ref{rfig3}).
During a given time interval $T$, $N$ targets
occurred and $N_F$ of them were not predicted.
The number of declared alarms was $A$,
with $A_F$ of them being false alarms.
The total  duration of alarms was $D$.
The error diagram shows the trade-off between the relative
duration of alarms $\tau=D/T$, the
fraction of failures to predict  $n=N_F/N$, and
the fraction of false alarms $f=A_F/A$.
In the $(n,\tau)$-plane the straight line $n+\tau=1$
corresponds to a random binomial  prediction --- at each step in
time  the alarm is declared with some
probability $\tau$ and  not declared with
probability $1-\tau$.
Given a particular prediction that depends on
our three parameters $(s,N_0,\Delta)$,
different points in the error diagram correspond to
different values of these parameters.
Error diagrams thus tally the score of a prediction algorithm's
successes and errors.
This score depends on the algorithm's adjustable parameters.
For example, raising the threshold $N_0$ will reduce the number
of alarms $A$ but may increase the number $N_F$ of failures to predict.
Raising $\Delta$, on the other hand, will increase the duration
alarms $D$ but may reduce the number of failures to predict $N_F$, etc.
A prediction algorithm is useful if:
(i) the prediction quality is better
than that of a random one, i.e.
the points on error diagram are close
to the origin and distant from the diagonal
$n+\tau=1$; and
(ii) this quality is fairly insensitive
to changes in the parameters.

\begin{figure}
\resizebox{6cm}{5.cm}{\includegraphics[width=7cm]{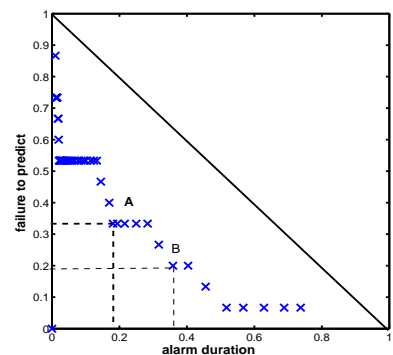}}
\caption{Error diagram: fraction of failures to predict as a
function of alarm duration.
The diagonal line correspond to a random prediction.
Deviations from this line depict predictive power of
the precursor.}
\label{rfig4a}
\end{figure}

\section{Results and Discussion}

\begin{figure}

 \resizebox{4.8cm}{4.8cm}{\includegraphics{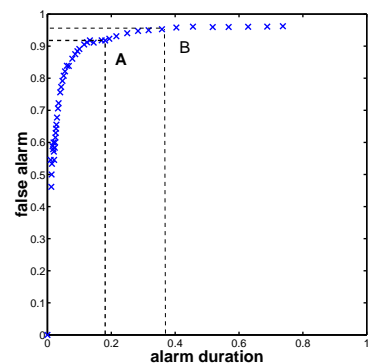}}

\caption{Error diagram: fraction of false alarm as a function of
alarm duration.The point at 20\% of alarm rate as
deduced from Figure (\ref{rfig4a}) correspond to a 90\% false alarm rate.
}
\label{rfig4b}

\end{figure}

We estimate the time predictability of volcanic eruptions based on
the increase of the daily seismicity rate.
The parameters of the algorithm are varied as follows:
$1<s<30$ days, $1\le N_0\le 100$ events per {\it s} days,
$1\le \Delta\le 30$ days.
The 30 day limit is the minimum time between two eruptions during
1988-2001.
The best predictions are obtained when averaging
seismicity rate over a 5 day window and
declaring an alarm for 5 days.
The predictive skills of our prediction scheme
are illustrated by the error diagrams of
Figures (\ref{rfig4a}, \ref{rfig4b}).
Each point in the error diagram corresponds to different values of
the threshold $N_0$ ranging from 1 to 100 events per 5 days,
other parameters are fixed as $s=5$ days, $\Delta=5$ days.
Error diagrams outline the whole range of possible
prediction outcomes; thus they are more convenient for
decision making than performance of ``the best'' single
version of prediction.
We observe for instance (Point A) that 65\% of the PdlF
eruptions can be predicted with 20\% of the time covered by alarms.
These results are of the same quality as that
obtained on the Etna or the Hawaii volcanoes.
For instance, using regional seismicity in a 120 km radius from
the Etna volcano, 50\% of the eruptions could
have been predicted within 40 days in the
1974-1990 period, which can be sorted as 80 \% of
the 11 flank eruptions, and no summit eruptions
[{\it Mulargia et al.}, 1991, 1992].
Decreasing the threshold $N_0$ yields an alternative prediction strategy
that favors a lower failure to predict rate and accepts a higher
alarm duration rate;
it is shown as point B on Figure (\ref{rfig4a}).
The choice of a particular prediction strategy must be always based on
the analysis of the entire error diagram;
different prediction strategies may be used in parallel to
complement each other (see more in [{\it Molchan,} 1997;
{\it Zaliapin et al.,} 2003]).

It is worth noticing that the performance of our simple prediction
algorithm, which is based on mere averaging of the seismicity rate,
is close to the performance of much more sophisticated algorithms that
use numerous seismic parameters to predict large observed
earthquakes [e.g. {\it Kossobokov et al.,} 1999].
The significant predictability we obtain here is still concomitant
of a fraction of false alarm larger than 90\% (Figure \ref{rfig4b}).
Because this predictability emerges from the use of
a daily seismicity rate only, we expect that a modification of the
above prediction strategy to include
earthquake location and magnitudes with deformation and geochemistry data
will improve this first quantitative analysis of eruption
prediction on PdlF.

\vskip 0.5cm

\acknowledgments
We gratefully thank OVPF staff in charge of the PdlF seismic network since 1980.
We thank A. Helmstetter, W. Z. Zhou, J. El-Khoury,
  T. Gilbert, D. Shatto, M. Collombet
and the ESS/UCLA seismo group for stimulating discussion.
We benefited from Professor V. Keilis-Borok's
lectures on time series analysis and pattern recognition during
the Spring 2003 Quarter at ESS/UCLA.
JRG is partially supported by  EC E-ruption project and EC EVG-CT-2001-00040,
Volcalert project.
IZ is partly supported by INTAS, grant 0748.

  \end{article}

\end{document}